# Spontaneous back-pain alters randomness in functional connections in large scale brain networks: A random matrix perspective

Gurpreet S. Matharoo[1,2]* and Javeria A. Hashmi[3]†

[1] ACENET, St. Francis Xavier University, Antigonish, NS, B2G 2W5, Canada.
[2] Department of Physics, St. Francis Xavier University, Antigonish, NS, B2G 2W5, Canada.
[3] Department of Anesthesia, Pain Management and Perioperative Medicine, Dalhousie University, Halifax, NS, Canada.

* gmatharo@stfx.ca (Corresponding author)
† javeria.hashmi@dal.ca

## Abstract

We use randomness as a measure to assess the impact of evoked pain on brain networks. Randomness is defined here as the intrinsic correlations that exist between different brain regions when the brain is in a task-free state. We use fMRI data of three brain states in a set of back pain patients monitored over a period of 6 months. We find that randomness in the task-free state closely follows the predictions of Gaussian orthogonal ensemble of random matrices. However, the randomness decreases when the brain is engaged in attending to painful inputs in patients suffering with early stages of back pain. A persistence of this pattern is observed in the patients that develop chronic back pain, while the patients who recover from pain after six months, the randomness no longer varies with the pain task. The study demonstrates the effectiveness of random matrix theory in differentiating between resting state and two distinct task states within the same patient. Further, it demonstrates that random matrix theory is effective in measuring systematic changes occurring in functional connectivity and offers new insights on how acute and chronic pain are processed in the brain at a network level.

## 1. Introduction

Chronic pain represents a major clinical, social, and economic problem for societies worldwide. The principal complaint is of unremitting physical pain that does not abate with standard analgesics [1–3]. The experience of pain is quite different across the population and persists for different durations between individuals. Pain is in essence a threat signal that we localize to a part of the body in the form of an unpleasant sensation. This sensation accompanies a strong negative emotion that works as an aversive signal which is necessary for learning proper avoidance behaviors. In some people, this signal becomes accentuated and tends to persist for long periods of times extending over months to years. These individuals very often show no signs of tissue damage or underlying pathology in the site where they are feeling pain. Brain imaging studies suggest that chronic pain alters the nervous system so that the brain perceives persistent pain due to maladaptive processes in the brain. An expedient approach for understanding these maladaptive processes is to observe how back pain transitions to a chronic form.

Thus, we know that in some patients, persistent back pain is acute and persists for a few weeks to be classified as subacute back pain (or SBP). This early stage of persistent back pain remits in some individuals, while for others, it persists for months to years and this enduring back pain is classified as chronic (Chronic Back Pain or CBP). Brain responses to back pain have been reported to change over time as people with subacute back pain develop chronic back pain. While any initial instance of self-report of spontaneous occurrence of back pain activates brain regions such as the insula and the anterior cingulate cortex that customarily respond to acutely evoked pain, over time, these instances correspond with activations in regions that process fear (amygdala) and self-referential thinking (medial prefrontal cortex). In a recent longitudinal study [3], it has been clearly demonstrated (with pictorial representations) that persistence of back pain alters brain responses. A large cohort of people with CBP, it was established that brain connectivity is also altered by persistent pain, where regions with the highest connectivity (hubs) show a deviation in their pattern across the brain relative to healthy controls and shows increases in modularity in sensory areas of the brain [4].

The reasons and neural mechanisms due to which back pain transitions from subacute to chronic are still ambiguous, and the pursuit to find neurological reasons for this transition is central to contemporary pain research. In recent years, there have been successful attempts in relating CBP to specific brain activity [5] whereby neuroimaging method of functional Magnetic Resonance Imaging (fMRI) is used to study the correlations between CBP and brain activity. fMRI makes use of the fact that neuronal activity is partly coupled with increases in blood flow in the observed parts of the brain and it images these changes as a haemodynamic response to brain activity. This particular form of fMRI is also referred to as blood-oxygenation-level-dependent (BOLD) fMRI and it offers high spatial resolution. A useful adaptation of this approach is to measure how slow temporal fluctuations (0.01-0.15 HZ) are between different brain regions and this statistical dependency is referred to, more generally, as functional connectivity. Identification of functional networks from fMRI data has gained importance in the recent years as it provides critical information about correlations between different regions of brain, and how these correlations are affected in various conditions [6,7]. The network properties that emerge from large-scale correlations have been shown to be altered in neuropsychiatric and chronic conditions such as CBP[5,8–12]. It is still a challenge to understand the dynamic transition of brain between different states as a result of back-pain. It is because brain is a fairly complex system whereby neurons are constantly interacting with each other often resulting in higher brain functions [13,14] and in the formation of functional networks, even in the absence of any stimuli. Though large-scale functional connectivity is often studied using clustering techniques or principles of graph theory[15], there is a need to apply the concepts and methodologies developed in the context of the theory of random matrices for observing systematic transitions in brain states.

Random Matrix Theory (RMT) was originally developed in the nuclear physics applications, where nuclei can have many possible states and energy levels and, and their interactions are too complex to be described accurately. In such a scenario, one settles for a model that captures the statistical properties of the energy spectrum. RMT finds extensive applications in the statistical studies of various complex systems such as quantum chaotic systems, complex nuclei, atoms, molecules, disordered mesoscopic systems [16–24], atmosphere [25], financial applications [26], complex networks [27], societal networks [28], network forming systems [29,30], amorphous clusters [31–34], biological networks [35], protein networks [36,37], and cancer networks [38] etc. In recent years, RMT has also been applied towards brain network studies in studying universal behavior of brain functional connectivity and has been effective in detecting the differences in resting state and visual stimulation state[39,40]. Recently, attempts using RMT have also been made in brain functional network studies on attention deficit hyperactivity disorder (ADHD) [41].

RMT makes use of the fact that true information of the system is contained in the eigenvalues of a correlation matrix. Specifically, for brain networks, the eigenvalues represent the level of functional connectivity between different regions of interest (ROIs) in brain, and larger eigenvalues contain information about significant correlations (or strong connectivity), and therefore, about processes in brain. Recent studies have shown that ROIs in brain are correlated. Furthermore, these correlations closely follow the predictions of Gaussian Orthogonal Ensemble (GOE) of random matrices when the brain is in a state of rest (fully conscious). The clearest indication so far has come from EEG data[39], which further attributes the observed deviation from GOE predictions to visual stimulation; that is, true information. Other recent studies[40,41] also point to similar information, however, the overall findings are unclear. We hereby propose a hypothesis where, we refer to these observed correlations as random correlations, or in general, randomness, that exists at any given instant in brain network. When the brain is engaged in a task, this randomness would be expected to decrease, as brain regions would be connected in a coherent fashion relative to a task-free or resting state. These random correlations reach their normal levels at resting state. Thus, RMT may offer a principled approach for measuring systematic changes in randomness that occur in brain networks during perception and cognition.

Here we investigate whether the brain demonstrates a greater deviation from GOE predictions when it is engaged in detecting threats or experiencing discomfort from pain relative to perception of innocuous stimuli. Since the ability to properly detect and perceive pain is fundamental for survival, attending to pain can be expected to add systematic changes in brain connectivity and thus reduce random correlations in brain networks. On the other hand, maladaptive processing of pain inputs during a chronic stage of back pain may show a different behavior, relative to the SBP state. The ability to distinguish these two states using an integrative approach such as RMT could be useful for improving chronic pain diagnosis and prognosis and also for understanding the abnormalities in brain properties that contribute to CBP.

## 2. Materials and methods

In the following sub-sections, we describe the methodology and the workflow that we have followed for the present work in a chronological order:

### 2.1 Subject Classification

For the present work, we use fMRI data available on the open access data sharing platform for brain imaging studies of human pain (www.openpain.org). The complete dataset is a part of 5-year longitudinal study of transition to chronic back pain in which 120 patients were recruited initially.

At each visit, fMRI scans and McGill Pain Questionnaire Visual Analogue Scale (MPQVAS) measures were recorded for all the patients.

For the present RMT-based study, we use fMRI scans obtained from two visits namely, an initial visit where all patients report back-pain, and a follow-up visit six months after the initial visit, whereby some patients report remission of back-pain and others report persistence of back-pain. As a result, at the follow-up visit, based on the difference of MPQVAS measures for the two visits, the patients are classified in two groups. For group of patients whose MPQVAS values decrease by 30% or more than the corresponding value at initial visit, we classify them as "SBP recovering (or simply, recovering)" group, and the rest as "SBP persistent (or simply, persistent)" group. A pictorial representation of this classification is illustrated in Figure 1.

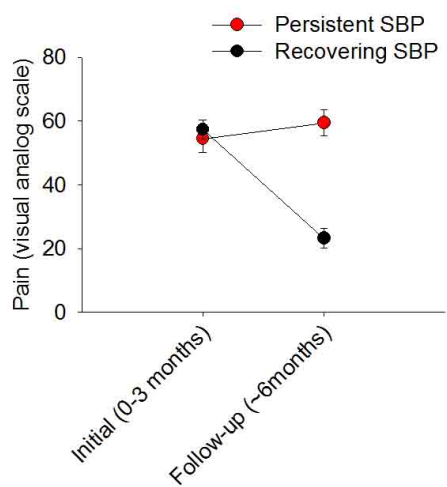

**Fig. 1:** Recovering SBP group in contrast to Persistent SBP group based on MPQVAS ratings. Each of the points denotes the mean value for the group. The error bars represent standard error of mean.

## 2.2 fMRI Tasks

All the participants were trained to perform two tasks using finger-span device with which they provided continuous pain ratings[3,5]. This device consisted of a potentiometer in which voltage was digitized. During the brain imaging sessions, the device was synchronized and time-stamped with fMRI image acquisition and connected to a computer providing visual feedback of the pain ratings [42]. We use data acquired from three different states:

**a) Resting State (RS):** A state of rest in which the participants are not thinking about any one thing in particular.

**b) Spontaneous Pain (SP):** A state of focusing and rating spontaneous changes in back pain. Here, the individuals saw a bar that increased or decreased in height on the y-axis scale (0-100). By changing the distance between the thumb and index finger, they could increase or decrease the height based on the intensity of pain they felt in their back on the scale. These measurements were recorded in real time and individuals continuously rated their back pain during the length of the entire brain scan.

**c) Standard Visual (SV):** A control state in which they are rating changes in length of a visual bar. Here, participants no longer rated their pain, instead they increased or decreased the distance between their fingers so that it matched the changes in the height of the bar on the scales y-axis. Thus, the SV condition represents a control condition that was unrelated to pain and only represents a visual-motor control task.

### 2.3 MRI data acquisition

The data for all participants and visits was collected by a 3T Siemens scanner. At first, MPRAGE type $T_1$ anatomical brain images were acquired followed by fMRI scans on the same day with the following parameter details [3]: EPI sequence: voxel size 1 X 1 x1 MM, Repetition time=2500MS; Echo Time=3.36MS; Flip angle = 9 degrees; In-Plane matrix resolution 256 X 256; slices 160, filed of view, 256mm. Functional MRI scans were acquired on the same day as the T1 scan and McGill Pain Questionnaire Visual Analogue Scale (MPQVAS) measures: multi-slice T2*-weighted EPI images with repetition time=2.5s, echo time=30ms, flip angle =90 degree, number of volumes =244, slice thickness =3mm, in-plane resolution =64 x 64.

### 2.4 Pre-processing of fMRI data

We use Freesurfer, FMRIB Software Library (FSL) v5.0, and Analysis of Functional Neuro-Images (AFNI) software to preprocess the data similar to procedures adapted for the 1000 Functional Connectomes project[43]. Data were slice time corrected, motion corrected, temporally band-pass filtered, and then further filtered to remove linear and quadratic trends using AFNI. Complete details of the preprocessing procedure are given in[44]. The registration was performed using FMRIB's Linear and non LINEAR Image Registration Tools for transformations from native functional and structural space to the Montreal Neurological Institute MNI152 template with 2 x 2 x 2 resolution, with further details given in[44].

### 2.5 Anatomical parcellation and construction of correlation matrix

The brain is anatomically parcellated by *an optimization of the Harvard/Oxford parcellation scheme* (OHOPS)[45]. In this scheme, the anatomical partitioning of cingulate, medial and lateral prefrontal cortices of Harvard Oxford Atlas was increased and in addition, anatomical partitioning of insular label was also performed, and the single Region of Interest (ROI) spanning the entire insula in Harvard Oxford Atlas was further subdivided based on a previous scheme[46]. The complete OHOPS consisted of a total of 131 regions[45]. Each ROI was designated as a node and the BOLD time series were extracted from each node and averaged to generate 131 time series for each subject. Following this, the whole brain networks were constructed, and network measures were assessed using the Brain Connectivity Toolbox, with formulae used for calculating network measures described in[47]. The brain networks are usually assortative in nature[48,49].

For each patient, the BOLD time series in each region was correlated with every other region to create a 131 x 131 symmetric correlation matrix based on Pearson's correlation coefficients given by:

$$corr(X, Y) = \frac{cov(X, Y)}{\sigma_X \sigma_Y}$$

or, which can also be written as:

$$corr(X, Y) = \frac{\sum_{i=1}^{n}(x_i - \overline{x})(y_i - \overline{y})}{(n-1)\sqrt{\frac{\sum_{i=1}^{n} x_i^2 - n\overline{x}}{n-1}} \sqrt{\frac{\sum_{i=1}^{n} y_i^2 - n\overline{y}}{n-1}}}$$

Here, X and Y are two distinct time series, each made up of $n$ time points, $x_i$ and $y_i$ respectively. For the present case, there are 240 time points ($n = 240$) for each time series. $\overline{x}$ and $\overline{y}$ are the respective means for two time series (x and y). By definition, the diagonal elements of the matrix are 1, as it represents self-correlation and the off-diagonal elements result in a symmetric matrix. Such correlation matrices are not only symmetric, but they are also positive semi-definite[50,51], with all eigenvalues being non-negative. This correlation matrix is then diagonalized and eigenvalues ($\lambda$) are obtained. In the present case, there are 131 eigenvalues, few eigenvalues are zeros, and remaining have positive values. It must be remembered that not all ROIs are a part of active brain network at a given time and hence, very small eigenvalues are usually ignored, and the related correlations are unimportant from functional connectivity perspective. In the present cases, usually the first 40 (around 30%) eigenvalues are extremely small from computational perspective. Hence, we leave them out from the subsequent analysis.

### 2.6 Unfolding of data

Fluctuations around the eigenvalue spectra are studied using standard methods of RMT. The first step is to unfold the data, meaning, the eigenvalues are arranged in an increasing (cumulative) order and are then mapped using an analytical function in such a way that the average spacing between two successive eigenvalues is unity. This ensures all the eigenvalues are on same footing. The analytical fitting function used for unfolding need not be unique and, is generally different for different systems[30–34]. For this study, the eigenvalue spectra of all the correlation matrices generated is approximated extremely well by a function of the form

$$(a - b * e^{-c\lambda^{1/d}})$$

where a, b, c, and d, are best-fit parameters and $\lambda$ is the eigenvalue. Figure 2 shows a plot of the cumulative eigenvalue density along with the analytical fitting function. We leave out a small portion of eigenvalues at the upper end (3 or 4 eigenvalues) in order to achieve the best fit, something which has been a standard practice in other works [30–34]. We deal with unfolded eigenvalues from this point onwards.

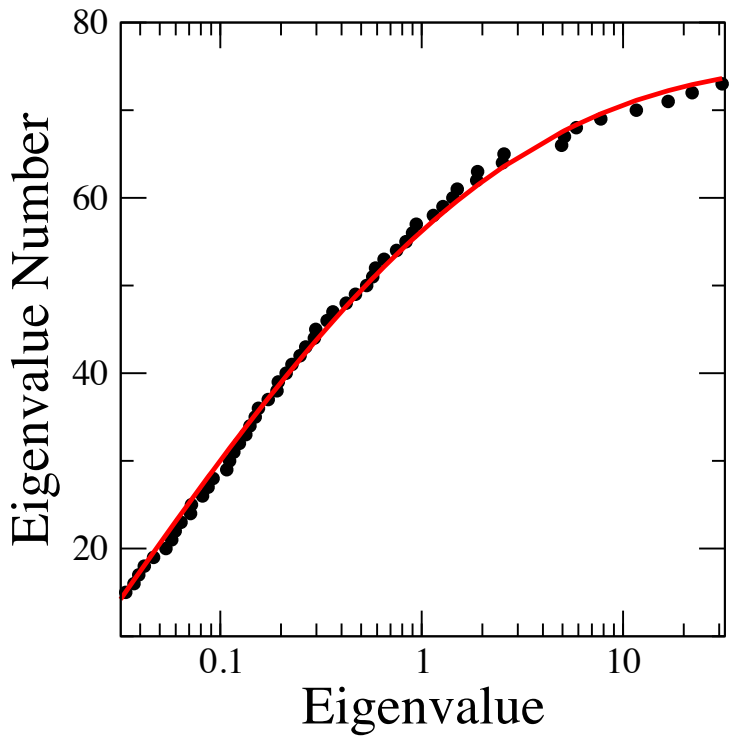

**Fig. 2:** Eigenvalue number vs eigenvalue ($\lambda$) for a typical spectrum. Filled circles (black): Data. Continuous line (red): The best-fit using the functional form described in text.

## 3. RESULTS

We report the spectral statistics fluctuation properties of the eigenvalue spectra in the three brain states in individuals who were suffering with SBP (back pain for < 3 months). We also track what these properties looked like after 6 months in the group of individuals with SBP with persisting back pain[3,5,10,52]. Patients had all been pain free for one year prior to their subacute pain episode and had no history of any mental illness including depression. The individual details of patients are also available online on the data sharing platform. It must also be stated that none of the data from available subjects was excluded from the analysis.

### 3.1 INITIAL VISIT

For the initial visit, where all patients report back-pain, 68 SP and 62 SV scans are available. In addition, there are 27 RS scans available. Analysis of randomly picked individual eigenvalue spectra indicate that brain-states have fluctuation properties associated with the Gaussian orthogonal ensemble (GOE) of random matrices. To improve statistics, we combine information from all unfolded data. Figure 3a shows the normalized nearest-neighbor spacing distribution (NNSD) [p(s)] for RS, SP, and SV scans for the initial visit. Here, $s$ is the eigenvalue spacing. Superimposed is the GOE result, which is also approximated by Wigner's surmise as:

$$p(s) = \left(\frac{\pi s}{2}\right) * e^{-\pi s^2/4}$$

For all the cases, we find a good agreement with GOE. For RS scans, this is not really surprising. Here, the patients have been directed to remain awake and not to think on any one thing in particular. In such a scenario, we would expect maximum randomness, hence NNSD would agree with GOE. The agreement of SP and SV scans with GOE is however, a more interesting case. In SP scans, as the patients are focusing on their back-pain and simultaneously reporting the pain rating through the finger device, a lot of brain regions are expected to be involved in this task. As a result, if there were to be a deviation from the GOE, we would expect it to be in SP scans. However, we do not see any deviation of NNSD from the GOE results. Lastly, SV being a visual task, is an intermediate of RS and SP states. Here, patients are following a displayed visual while performing the finger-spanning task without specifically focusing on the back-pain, and once again, we find an excellent fit of NNSD with the GOE. A single-valued indicator that follows the p(s) function is the variance of nearest-neighbor spacing. We find this number between 0.297 and 0.320 for all the cases, which is quite close to 0.286, the number for GOE[31–33]. This agreement could be explained due to the fact that NNSD captures the correlations that exists between successive eigenvalues and does not have information about the long-range correlations. Short-ranged correlations, especially between the nearest-neighbors are quite strong, and hence not altered substantially by both, visual (SV) and pain-rating (SP) tasks. This result is also consistent to other brain-network studies[39–41,49] and hence, further strengthens the belief that there exists strong, stimuli-resistant random correlations between nearest-neighbors in the brain network.

Next, we take a look at the long-range (or higher order) random correlations. For this, we measure $\Sigma^2(r)$, the variance of the number of levels $n(r)$ within an interval of length $r$. The theoretical result for GOE is:

$$\Sigma^2(r) = \frac{2}{\pi^2}\left(ln(2\pi r) + 1.5772 - \frac{\pi^2}{8}\right)$$

The number variance is quite sensitive to changes, and is extremely sensitive to small systematic errors in the approximation to the analytical function used during unfolding[31,32]. Contribution of any such error to $\Sigma^2(r)$ grows as $r^2$, whereas the GOE prediction for $\Sigma^2(r)$ grows as $ln(r)$[34]. In Figure 3b, we plot $\Sigma^2(r)$ for RS, SP, and SV scans along with GOE and Poisson [$\Sigma^2(r) = r$] distributions for the initial visit. We observe that RS agrees with the GOE prediction over greatest domain, whereas we do see deviations for SV and SP scans with SP scans showing maximum deviation. This deviation is attributed to the relative tasks the subjects are performing in each case, with the pain-rating task showing maximum deviation. While it is on the expected lines to observe the variance agreement for RS scans to the GOE, it further demonstrates the efficacy of the RMT of capturing strong random long-ranged correlations when the brain is in a state of rest. We see a clear deviation from GOE for SP scans, whereby the patients are performing a pain-rating task. As stated before, a lot of brain regions are expected to participate in this task and, as a result, we see a clear decrease in randomness for SP scans. SV scans, however, present an interesting picture. We observe SV scans to show a good agreement with GOE for a greater range than SP scans, whereby the agreement matches with RS scans. While we do observe deviations from GOE eventually, the deviations are always less than SP deviations. This could be explained due to the nature of the task performed for SV scans. As the patients are not focusing on back-pain, the task involves only visual cortex to take part. In other words, compared to SP, this is an easier task to perform and, the difference between SV state and RS is quite subtle. As a result, fewer brain regions are expected to participate here. This inference is also consistent with the earlier results, whereby it is shown that salient percepts like pain engage more brain regions than visual stimulation[53–55]. This observed difference between the SP and SV scans is also the impact that SBP has on the brain networks. Additionally, also important here is the fact that RMT is able to capture the differences between two distinct task states.

### 3.2 FOLLOW-UP VISIT

At follow-up visit, which was approximately 6 months after the initial visit, the patients were made to repeat the same tasks (RS, SP and SV) and the corresponding scans were recorded. At this follow-up visit, while some patients recovered from persistent back-pain as a result of spontaneous remission of the condition (SBP recovering group), others experienced a persistence in their back-pain, and they represent the group who have developed chronic back-pain (persistent group). To define SBP persistent group, we separate participants with pain persisting for 6 months from those that recovered (SBP recovering) based on self-report of pain ratings observed using McGill Pain Questionnaire Visual Analogue Scale (MPQVAS). We compare the MPQVAS values at initial and follow-up visits. If the pain rating value of a particular subject decreases by 30% or more, the subject is classified as ``Recovering'', else, it is classified as ``Persistent'' (See Figure 1). Based on this classification, we have 18 RS, 17 SP, and 23 SV scans for Persistent group and 18 RS, 19 SP, and 17 SV scans for Recovering group.

Figure 4 shows NNSD for Persistent and Recovering groups. In both the plots, we observe the same trend for NNSD as it was at the initial visit. In both the plots, all the scans show an agreement with GOE predictions; an indicator of strong nearest-neighbor random correlations. The tasks at the follow-up visit are exactly same as the initial visit's tasks. As a result, we can state with a greater certainty that the NNSD captures short-ranged correlations effectively, and the randomness is undeterred by the pain stimuli.

We now take a look at the long-ranged random correlations. As mentioned before, this quantity is quite sensitive to the network changes that occur over a period of time. Figure 5 shows plots of $\Sigma^2(r)$ for Persistent and Recovering groups. In both the cases, we find RS scans staying close to GOE predictions. This once again is on expected lines. However, we find a striking difference between SP and SV scans in the two cases.

For the Persistent group, both SP and SV scans show deviations from the theory, with SP scans clearly showing greater deviations from theory, and SV scans showing only subtle deviations. The clear deviations of SP scans from GOE for the persistent group is also a reflection of the fact that they continue to experience the back-pain, hence, they are prone to chronic back-pain. And, as in the case of initial visit, the subtle differences between SV scans and theory could once again be attributed to the fact that visual stimulation task involves engagement of fewer brain regions. The recovering group, however, present a very interesting case. For the Recovering group, both SP and SV scans match GOE predictions over a larger domain and are indistinguishable from RS scans. Here, as a result of the medical treatment, the patients have experienced pain remission. As a result, they have none to very few pain events to report for SP scans. This observation once again demonstrates the efficacy of RMT in capturing the network changes in brain networks.

## 4. Conclusions and Discussion

Randomness is inherent in all brain networks and it follows the characteristics of GOE of random matrices. The resting state can be assumed as a normal state, and it defines normal or equilibrium levels of randomness. Both, cognitive tasks and salient percepts (for example, pain) decrease randomness as they require more brain regions to be focused. In network concept, resting state could also be assumed as more random state or a disordered state [27], and cognitive tasks and salient percepts force it to be more ordered. Hence, task-states can also be interpreted as more ordered than the normal state. Mathematically, it means deviations from the GOE predictions. Once the tasks are over, or the salient percepts are no longer there, we would expect the randomness to reach its normal or equilibrium levels.

For all the cases, our results demonstrate that the randomness shows maximum agreement with GOE for the RS scans and it decreases the most for SP scans. So, RS can be viewed as a most random state, and SP state can be viewed as a most-ordered state. SV state falls between the two. The resting state is important with regard to BOLD fMRI correlations, and the agreement with GOE could also be visualized as a single correlation structure that may adequately describe it [7]. Also, the continued agreement of the RS scans with GOE is also consistent with the reasoning that resting state BOLD correlations reflect processes concerned with long term stability of brain's functional organization, and generally do not reflect short term changes in cognitive content [7].

Further, our results demonstrate that RMT is able to differentiate between two different tasks within the same subject. Here, we find a pattern consistent with our hypothesis, with randomness decreasing when the brain is focused on attending to pain triggered in the back of their body. Here, GOE line represents maximum randomness and Poisson represents no randomness. However, due to the complexity of the experimental design, there could be many possible conjectures (including their combinations) explaining these observations.

First, as the patients are performing a pain-rating task, whereby they are focusing on the back-pain and reporting the ratings, the observed SP deviations could be attributed to back-pain. As it known from earlier studies that salient percepts such as pain are known to require more brain areas to be engaged than visual stimulation[53–55], we see an increased deviation for SP scans relative to SV scans in all the cases. As more brain regions are engaged in attending to pain, hence relative randomness between them decreases. At initial visit, all patients report back-pain, whereas at follow-up visit, only a subset of them report back-pain, and because their MPQVAS ratings demonstrate chronification of pain, the persistent group continues to experience back-pain over many months. Hence, this continued deviation of SP scans at the follow-up visit in the persisting CBP group could be viewed as a reflection of chronified pain that continues to affect the GOE pattern. It is also known that task states can alter the correlation structure of BOLD activity [7] and hence, the second possible conjecture is the saliency between the tasks themselves. While visual tasks are relatively easy to perform, pain-rating tasks could be much difficult as back-pain events are generally random. Hence, more attention is needed to perform these tasks, and thereby, we observe a decrease in randomness between the brain regions involved in these tasks.

Finally, in spite of the complexities in the experimental design in the present work, the observations presented here prepare a platform to study fMRI generated brain-networks using RMT. RMT could be effectively used in studying metastability of brain networks impacted by other neuro-psychiatric disorders. Clues from RMT studies on other physical systems, especially liquids and amorphous solids, could be useful here. For example, normal modes studies on liquids [29,30] and amorphous systems [27–30] have revealed universal properties whereby, the fluctuations around the mean spectral densities for stable configurations (local minima) follow GOE, and deviations from GOE are observed for non-stable configurations. In this context, RMT could be used in the energy landscape studies of brain in the detection of metastable states. An inherent shortcoming of this method is that it is statistical in nature. However, suitable modifications and adaptations of the methodology in artificial neural networks would be extremely helpful. The fact that the resting state is a state with maximum randomness could then be used as a key component in determining any systematic or mechanical errors in fMRI scans. Also, it could reflect on the long-term stability of brain's functional organization.

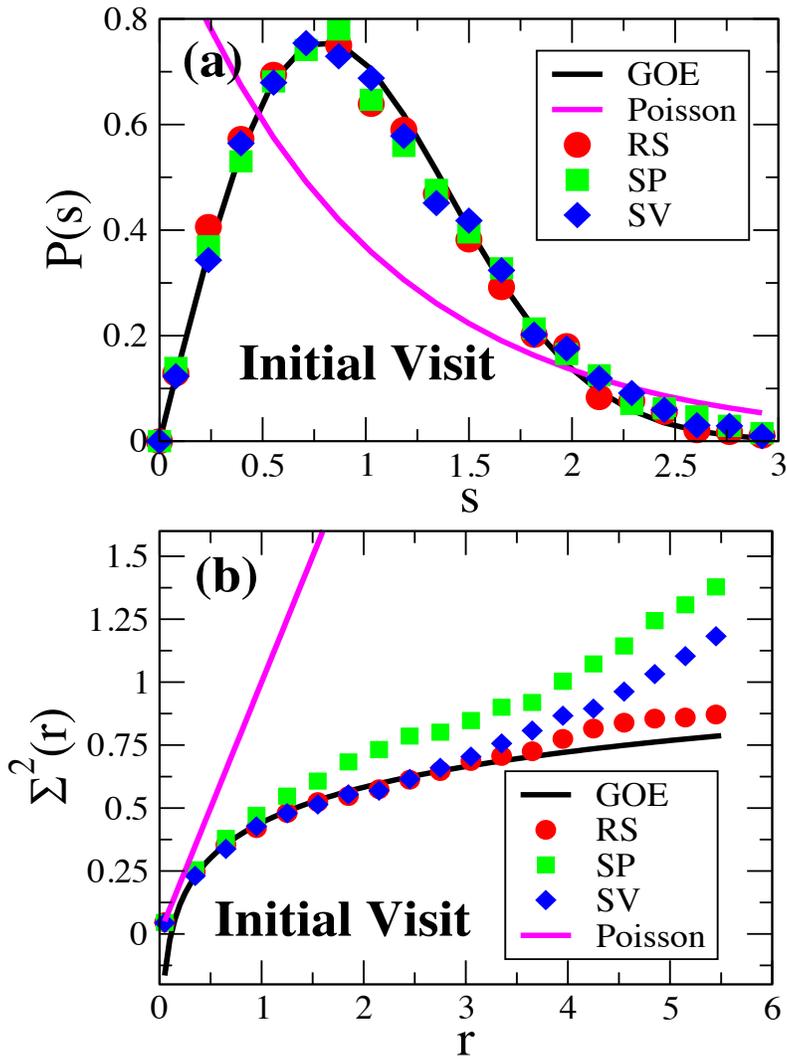

**Fig. 3:** **(a)** Normalized neighbor spacing *(s)* vs probability density *p(s)* for resting state (red circles), spontaneous pain (green squares), and standard visual (blue diamonds) scans for the initial visit. Black line represents GOE prediction and magenta line represents Poisson distribution; **(b)** Variance of the number of levels in intervals of length *r* shown as a function of *r* for resting state (red circles), spontaneous pain (green squares), and standard visual (blue diamonds) for the initial visit. Black line represents GOE prediction and magenta line represents Poisson distribution.

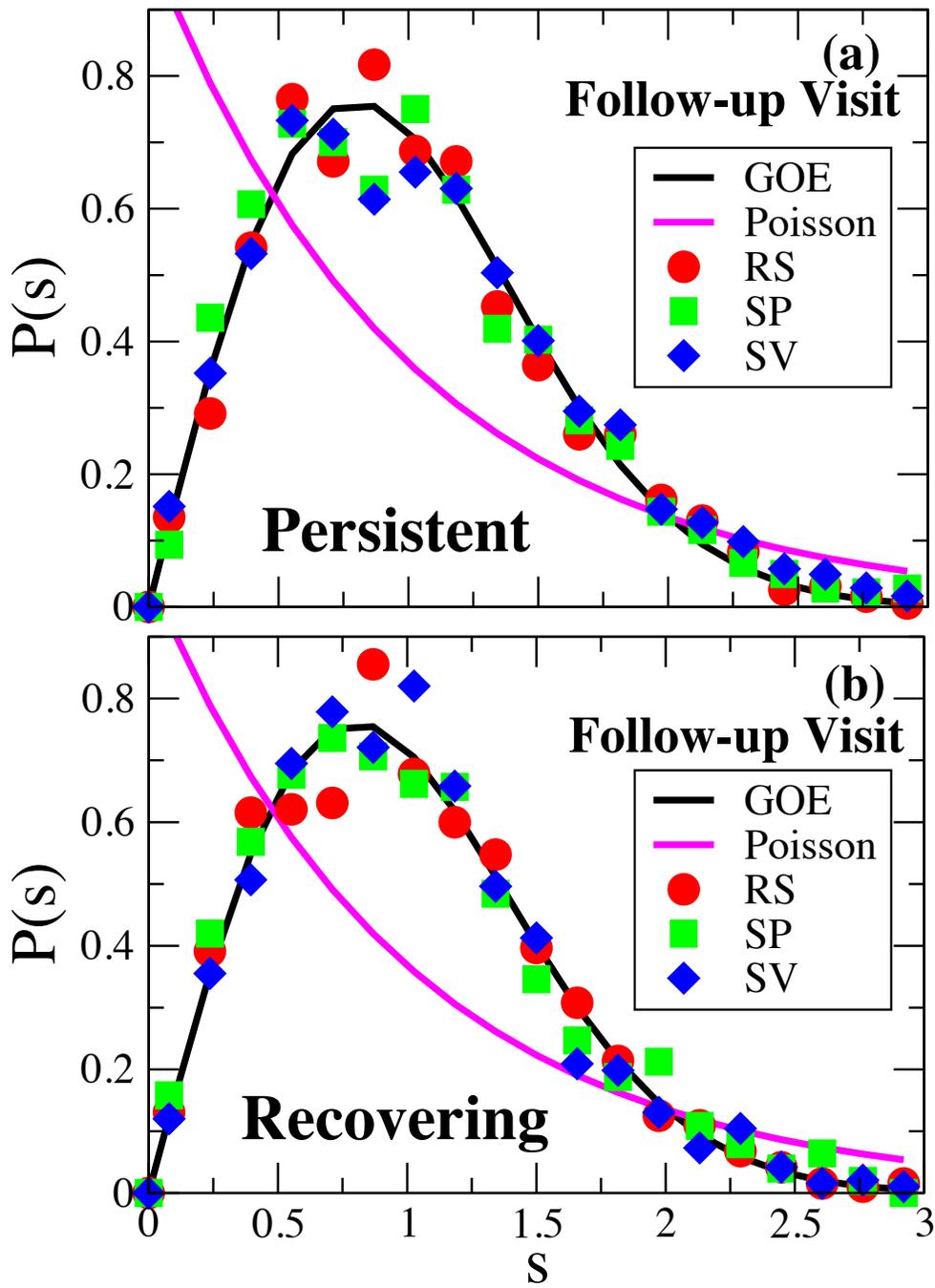

**Fig. 4:** Normalized neighbor spacing *(s)* vs probability density *p(s)* for resting state (red circles), spontaneous pain (green squares), and standard visual (blue diamonds) scans for **(a)** Persistent, and **(b)** Recovering groups in the follow-up visit. Black line represents GOE prediction and magenta line represents Poisson distribution.

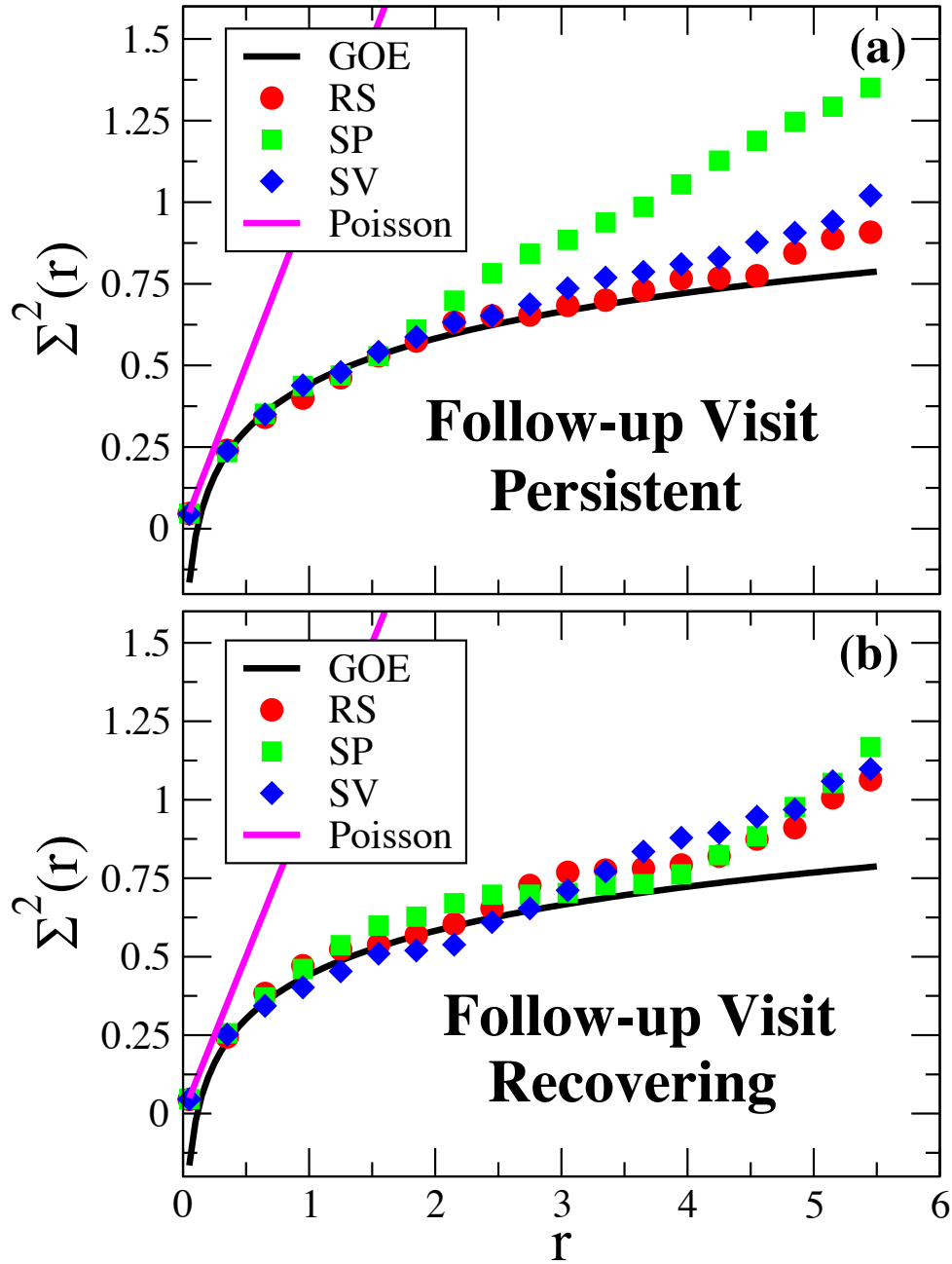

**Fig. 5:** Variance of the number of levels in intervals of length *r* shown as a function of *r* for resting state (red circles), spontaneous pain (green squares), and standard visual (blue diamonds) for **(a)** Persistent, and **(b)** Recovering groups in the follow-up visit. For both visits, black line represents GOE prediction and magenta line represents Poisson distribution.

## Availability of data and materials

Data used in the preparation of this work were obtained from the OpenPain Project (OPP) database (www.openpain.org). The OPP project (Principal Investigator: A. Vania Apkarian, Ph.D. at Northwestern University) is supported by the National Institute of Neurological Disorders and Stroke (NINDS) and National Institute of Drug Abuse (NIDA).

The preprocessing codes and the behavioral data (including MPQVAS data) used for the present work could be obtained by request from the authors.

**Acknowledgements:** GSM would like to thank Karl-Peter Marzlin for useful discussions and suggestions. We thank ACENET and Compute Canada for computational resources.

**Funding:** This research did not receive any specific grant from funding agencies in the public, commercial, or not-for-profit sectors.

**Conflict of Interest:** The authors declare no conflict of interest